\documentclass[usenatbib]{iau}
\usepackage{graphicx}
\usepackage{subfig}

\title[Quasar MIR Spectral Templates] 
{High Signal-to-Noise Ratio Mid-Infrared Quasar Spectral Templates}

\author[Allison~R.~Hill et al.]   
{Allison~R.~Hill$^1$, S.~C.~Gallagher$^1$, R.~P.~Deo$^1$,
  E.~Peeters$^{1,2}$ \and Gordon~T.~Richards$^{3,4}$}

\affiliation{$^1$The University of Western Ontario \\ 1151 Richmond
  Street, London, ON,  N6A3K7, Canada \\ email: {\tt ahill49@gmail.com} \\[\affilskip]
$^2$SETI Institute \\ 189 Bernardo Avenue, Suite 100, Mountain View,
CA 94043, USA\\[\affilskip]
$^3$Dept. of Physics \\ 3141 Chestnut Street, Drexel University,
Philadelphia, PA 19104, USA\\[\affilskip]
$^4$Max Planck Institut f\"{u}r Astronomie \\ K\"{o}nigstuhl 17,
Heidelberg, Germany 69117\\}

\pubyear{2013}
\volume{304}  
\pagerange{--}
\jname{Multiwavelength AGN Surveys and Studies}
\editors{A.C. Editor, B.D. Editor \& C.E. Editor, eds.}
\begin{document}

\maketitle

\begin{abstract}
  Mid-infrared (MIR) quasar spectra exhibit a suite of emission
  features including high ionization coronal lines from the narrow
  line region (NLR) illuminated by the ionizing continuum, and hot
  dust features from grains, as well as polycyclic aromatic
  hydrocarbons (PAH) features from star formation in the host
  galaxy. Few features are detected in most spectra because of
  typically low signal-to-noise ratio (S/N) data. By generating
  spectral composites in three different luminosity bins from over 180
  {\em Spitzer} Ifnfrared Spectrograph (IRS) observations, we boost
  the S/N and reveal important features in the complex spectra. We
  detect high-ionization, forbidden emission lines in all templates,
  PAH features in all but the most luminous objects, and broad
  silicate and graphite features in emission whose strength increases
  relative to the continuum with luminosity. We find that the
  intrinsic quasar spectrum for all luminosity templates is
  consistent, and the differences in the spectra can be explained by
  host galaxy contamination in the lower luminosity templates. We also
  posit that star formation may be active in most quasar host
  galaxies, but the spectral features of star formation are only
  detectable if the quasar is faint.
\end{abstract}

\firstsection 
\section{Introduction}

Quasars are among the most luminous phenomena in the universe,
generating substantial radiation across a wide range of the
electromagnetic spectrum. The typical quasar
spectral energy distribution has two distinct peaks, in the optical/UV
and the MIR (see figure 11 in \cite[Richards et al. 2006a]{ric06a}). The optical/UV peak (the
`big blue bump') is emission from the accretion disk on subparsec
scales around the supermassive black hole. The gas temperatures are
too high in this region for dust to survive. However, at distances
greater than approximately 1~pc from the supermassive black hole
(depending on the luminosity of the quasar), temperatures drop such
that dust is no longer sublimated. Beyond this point (the dust
sublimation radius), grains will be heated by emission from the
accretion disk and radiate in the MIR (\cite[Antonucci 1993]{ant93}). This thermal
emission is the source of the characteristic `MIR bump' in the quasar
SED.

Quasars have been studied extensively in the optical/UV, as relatively
unobscured quasars are readily studied from the ground in this
regime. As a result, the optical has the advantage of large
($>100~000$ quasar) sample sizes provided by surveys such as the Sloan
Digital Sky Survey (SDSS) (\cite[Schneider et al. 2010]{sch10}). Only since the advent of
space-based facilities such as {\it Spitzer} (\cite[Werner et al. 2004]{wer04}) has the
sensitivity been available to take spectra of similar objects in the
MIR. Even with its unprecedented sensitivity, these spectra typically
have low signal-to-noise ratios (S/Ns). Using an archival sample of
SDSS quasars observed with {\it Spitzer}, we constructed spectral
composites to boost the S/N, thus revealing faint features, to 
characterize typical quasar emission in the MIR.

\section{Quasar Sample Selection and Template Construction}

We coordinate cross-matched the SDSS quasar catalogue (\cite[Schneider et al. 2010]{sch10})
with the {\it Spitzer} Infrared Spectrograph (IRS; \cite[Houck et al. 2004]{hou04})
archive, using a match radius of 2$^{\prime\prime}$, yielding 184 low
resolution ($R\sim60$--130) MIR spectra (see Table 1 of
\cite[Hill et al. 2013]{hil13} for a summary of the sample). We selected all data
which had short-low and/or long-low modules (although not all objects
necessarily had both modules). Although our data span a wide range of
luminosities and redshifts, most of our objects occupy the low
luminosity, low redshift regime of the parameter space.  The details
of the data reduction are described in \cite[Hill et al. 2013]{hil13}.

\begin{figure}[b]
\begin{center}
 \includegraphics[trim=0cm 0cm 0.1cm 0cm, clip=true,width=\linewidth]{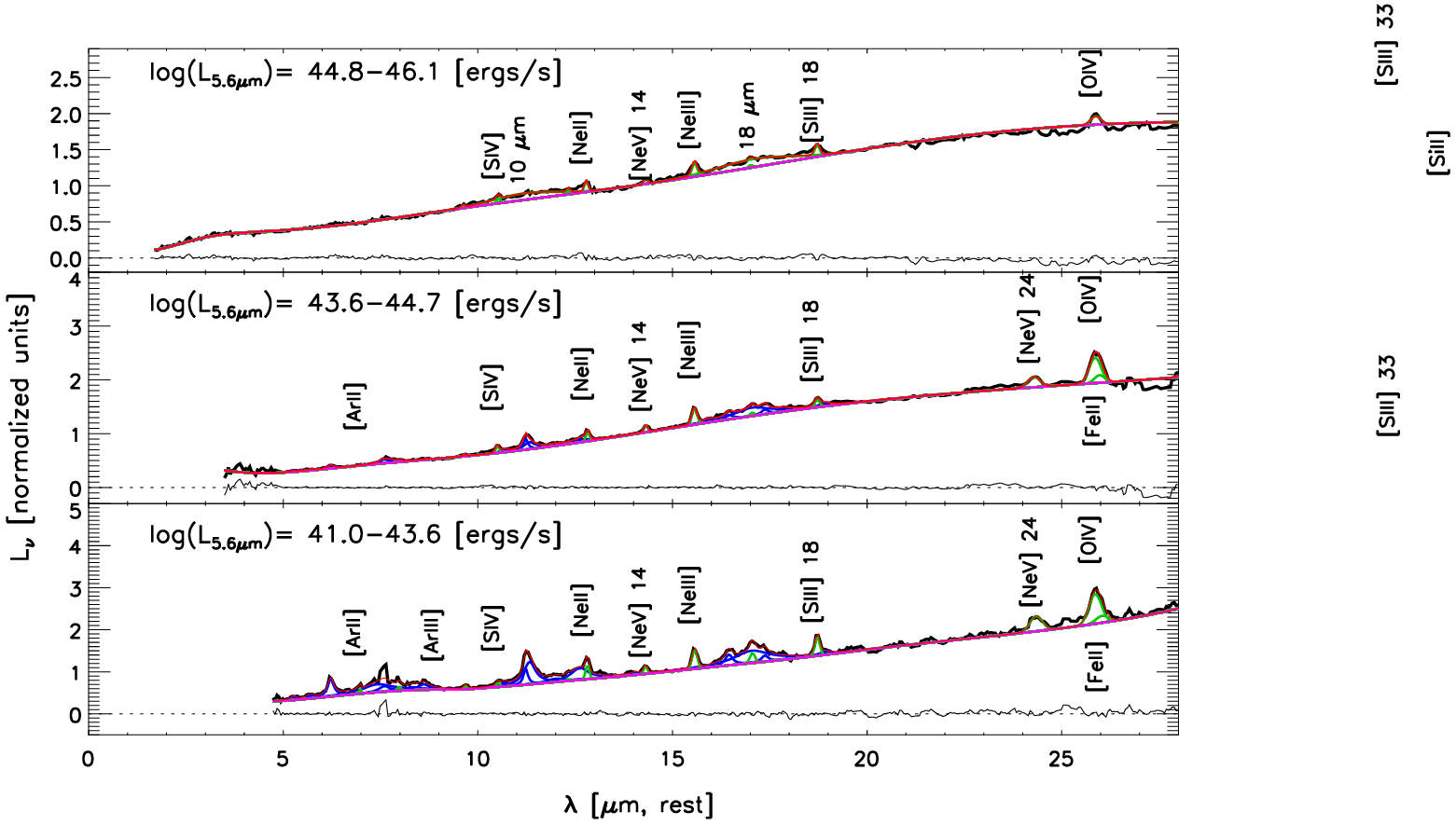} 
 \caption{Rest-frame spectral composites with PAHFIT models overplotted, sorted by decreasing luminosity from top to bottom. Model-subtracted residuals straddling the dotted black line at $L_{\nu}=0$.  Green and blue curves show the narrow line and PAH features respectively. Magenta is the continuum fit, and red is the total model. Note the absence of PAH features in the top panel; the 10 and 18~$\mu\/m$ silicate features are only evident in the top panel.}
   \label{fig:total}
\end{center}
\end{figure}

The objects were ordered according to their $5.6~\mu\/m$ continuum
luminosity, and divided equally into three bins of 61 objects per
bin. The $\log(L_{5.6~\mu\/m})$ ranges for each tertile are
41.0--43.6, 43.6--44.7, and 44.8--46.1 [erg/s]. For each bin, a
template spectrum was generated (Figure~\ref{fig:total}).  Also shown
in Figure~\ref{fig:total} are the {\verb PAHFIT} model fits for the
data.  {\verb PAHFIT} is a spectral decomposition code that separates
the MIR spectra into its PAH, narrow emission line, and continuum
components (\cite[Smith et al. 2007]{smi07}). Because {\verb PAHfit} was not designed with
AGN in mind, broad gaussian functions were added to fit the broad
emission features from silicates at $10$ and $18~\mu\/m$.

\section{Results and Discussion}

\subsection{Spectral Trends with Luminosity}

A visual inspection of Figure~\ref{fig:total} reveals a few
qualitative trends with increasing $5.6\mu\/m$ continuum luminosity:
the equivalent widths (EWs) of the forbidden lines decreases, the EWs
of the PAH features decreases until they are no longer detectable in
the most luminous tertile, and the continuum flattens towards longer
wavelengths. (For a more quantitative analysis of these trends, and a
more in-depth discussion of the narrow lines see \cite[Hill et al. 2013]{hil13}.)
Because our most luminous objects also tend to be at higher redshifts,
we are also able to see an emission bump blueward of $5~\mu\/m$ in the
most luminous tertile, which was best-fit in PAHFIT with a
black body of 750~K. The source of the emission blue-ward of
$5~\mu\/m$ is from dust with a higher sublimation temperature than
silicates, possibly due to graphites located closer to the central
engine than the silicate sublimation radius.

\subsection{Quasars and Their Host Galaxies}

The lack of PAH emission in many luminous quasar spectra is often
taken to indicate either low levels of star formation in their host
galaxies (e.g., \cite[Schweitzer et al. 2006]{sch06}) or that the hard
ionizing spectrum of the quasar is destroying the requisite molecules
(e.g., \cite[Genzel et al. 1998]{gen98}).  We instead investigate whether
the problem could be one of contrast and S/N, i.e., that the strong
MIR continuum of the quasar could be hiding vigorous star formation.

To determine the luminosity required for the strong PAH emission seen
in the least luminous template to be detectable in the most luminous
quasar template, we took the best-fitting PAH model shown in the
bottom panel of Figure~\ref{fig:total} and added gaussian noise. To
determine the appropriate PAH luminosity, we opted for the most
optimistic case: the PAH model was assumed to come from the highest
luminosity object in the lowest luminosity bin, and the observed
template was assigned the lowest luminosity of the objects in the
highest luminosity bin.  Even in our most optimistic case, there are
no prominent PAH features visible in the residual spectrum made by
subtracting the most luminous template model from the
template+PAH spectrum. In order for the $11.3~\mu\/m$ complex to be
detected at $3~\sigma$, the PAH model luminosity had to be scaled up
by a factor of 2.5. If we assume that PAH luminosity is correlated
with star formation rate (\cite[Schweitzer et al. 2006]{sch06}), this suggests there could be a
non-negligible amount of star formation activity in quasar host
galaxies; the spectral signatures of star formation are just not
visible if the quasar is too luminous.

\begin{figure}
	\centering
	\includegraphics[width=4 inches]{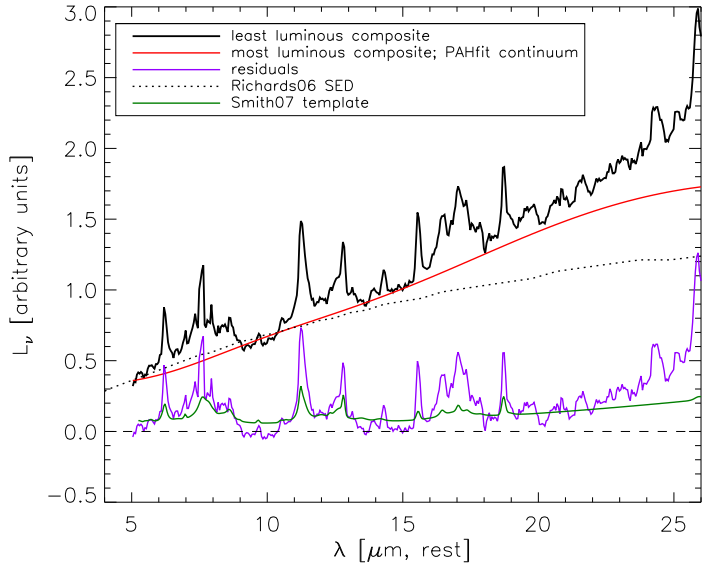}
	\caption{Comparison of the continuum of the most luminous
          template with the least luminous template.  The PAHFIT model
          continuum of the most luminous template (solid red line) was
          normalized to the least luminous template at $5~\mu\/m$
          (solid black line). The \cite[Richards et al. (2006a)]{ric06a} SED (dotted black
          line) is overplotted for comparison. The residuals from the
          subtraction of the PAHFIT model continuum (solid purple
          line) are above the 0.0 line (dashed black line). A
          starburst galaxy template from \cite[Smith et al. (2007)]{smi07} (solid green
          curve) is normalized to the $12.7~\mu\/m$ complex in the
          residuals. The features in the residuals can largely be
          accounted for by star formation in the host galaxy; this
          suggests that the underlying quasar continua are consistent
          for all luminosity templates.}
	\label{fig:cont}
\end{figure}

PAH emission is most prominent in the least luminous tertile, and
undetected in the most luminous tertile. This implies that host galaxy
contamination is strongest in the least luminous tertile, and the most
luminous tertile is more representative of the intrinsic MIR quasar
emission. The primary differences between the three templates are
therefore consistent with being caused almost entirely by different
levels of host galaxy contamination. To illustrate the host galaxy
contamination in the least luminous template, we normalized the
  PAHFIT model continuum from the most luminous template to the
$5~\mu\/m$ intensity in the least luminous template and subtracted
them (Figure~\ref{fig:cont}). Under the assumption that the continuum
from the most luminous template is more representative of the
intrinsic quasar continuum, the residuals should reflect the host
galaxy contribution. We over-plotted a galaxy starburst template from
\cite[Smith et al. (2007)]{smi07} (their template 3), and normalized the template to the
$12.7~\mu\/m$ complex of the residuals. The \cite[Smith et al. (2007)]{smi07} template
does a satisfactory job of reproducing many of the main features of
the spectrum as well as its overall shape. Given the diversity of
observed star-forming galaxy MIR spectra, we find this level of match
to be surprisingly good.

Considering the lack of PAH emission, and the results of
Figure~\ref{fig:cont}, we suspect that the most luminous template is
representative of the true MIR spectrum of a quasar. Furthermore,
quasar host galaxies may contain non-negligible levels of star
formation, the signatures of which are not detectable if the quasar is
too luminous.

\end{document}